\documentclass[showpacs,preprintnumbers,amsmath,amssymb,superscriptaddress]{revtex4}

\usepackage{txfonts}
\usepackage{amsfonts}
\usepackage{graphicx}
\usepackage{dcolumn}
\usepackage{bm}

\begin{document}

\preprint{Physcial Review E \textbf{78}, 051909 (2008).}

\title{Detection of subthreshold pulses in neurons with channel noise}

\author{Yong Chen}
\altaffiliation{Corresponding author. Email: ychen.ch@gmail.com}
\affiliation{Institute of Theoretical Physics, Lanzhou University, Lanzhou $730000$, China}
\affiliation{Key Laboratory for Magnetism and Magnetic materials of the Ministry of Education, Lanzhou University, Lanzhou $730000$, China}

\author{Lianchun Yu}
\affiliation{Institute of Theoretical Physics, Lanzhou University,
Lanzhou $730000$, China}

\author{Shao-Meng Qin}
\affiliation{Institute of Theoretical Physics, Lanzhou University,
Lanzhou $730000$, China}

\date{\today}

\begin{abstract}
Neurons are subject to various kinds of noise. In addition to synaptic noise, the stochastic opening and closing of ion channels represents an intrinsic source of noise that affects the signal processing properties of the neuron. In this paper, we studied the response of a stochastic Hodgkin-Huxley neuron to transient input subthreshold pulses. It was found that the average response time decreases but variance increases as the amplitude of channel noise increases. In the case of single pulse detection, we show that channel noise enables one neuron to detect the subthreshold signals and an optimal membrane area (or channel noise intensity) exists for a single neuron to achieve optimal performance. However, the detection ability of a single neuron is limited by large errors. Here, we test a simple neuronal network that can enhance the pulse detecting abilities of neurons and find dozens of neurons can perfectly detect subthreshold pulses. The phenomenon of intrinsic stochastic resonance is also found both at the level of single neurons and at the level of networks. At the network level, the detection ability of networks can be optimized for the number of neurons comprising the network.

\end{abstract}

\maketitle

\section{\label{sec:level1}INTRODUCTION}

It is well known that neurons are subject to various kinds of
noise. Intracellular recordings of cortical neurons \textit{in
vivo} consistently display highly complex and irregular
activity~\cite{Shink}, resulting from an intense and sustained
discharge of presynaptic neurons in the cortical network. Previous
studies have suggested that this tremendous synaptic activity, or
synaptic noise, may play a prominent role in neural information
transmission as well as in neural information
processing~\cite{Volgushev}. For example, with stochastic
resonance (SR), synaptic noise facilitates information transfer or
allows the transmission of the subthreshold
inputs~\cite{Gammaitoni}. Indeed, SR induced by synaptic noise has
been extensively studied in a single neuron and neural populations
both experimentally and
numerically~\cite{Stacey,William,William2}.

While the synaptic noise accounts for the majority of noise in
neural systems, another significant noise source is the stochastic
activity of ion channels. Voltage-gated ion channels in neuronal
membranes fluctuate randomly between different conformational
states due to thermal agitation. Fluctuations between conducting
and non-conducting states give rise to noisy membrane currents and
subthreshold voltage fluctuations. Recently, much effort has been
devoted to this field and channel noise is now understood to have
important effects on neuronal information processing capabilities.
Studies show that channel noise alters action potential dynamics,
enhances signal detection, alters spike-timing reliability, and
affects the tuning properties of the
cell~\cite{White,Schneidman,Schneidman2,Schreiber} (for review
see~\cite{White2}).

Detection of small signals is particularly important for animal survival~\cite{Svirskis}.
Both experimental and numerical studies have found, as depicted by SR, that synaptic noise
can enhance the detection of subthreshold signals in nonlinear and threshold-detecting systems.
For channel noise, there have been many papers concentrating on SR induced by channel noise~\cite{Adair, Schmid},
and their results suggest that neurons may utilize channel noise to process subthreshold signals. However,
it is still unclear whether reliable detection of subthreshold signals could obtained for single neuron
if neurons do utilize SR to
process signals. On the other hand, as a intrinsic noise source of neurons, channel noise is mostly studied within
single neurons. Since recent studies suggest that channel noise enhances synchronization of two coupled neurons\cite{Yu},
it is natural to ask whether channel noise could take effects in the network level.

In this study we focus on subthreshold pulse detection in neurons
with channel noise. First, using the stochastic Hodgkin-Huxley
(SHH) neuron model, we study the effects of channel noise on the
response properties of a single neuron to subthreshold pulse
input. We find that a SHH neuron fires spikes a higher than
average level in response to a subthreshold stimulus. The average
response time decreases while the variance increases as the
channel noise amplitude increases. This result is explained well
by the phase plane analysis method~\cite{Rinzel}. Then, we
evaluate the subthreshold signal detection ability of a SHH neuron
under the pulse detection scenario proposed by Wenning et
al.~\cite{Gregor}. They reported that colored synaptic noise can
enhance the detection of a subthreshold input. However, since the
total error is always greater than $0.5$, they argued that
biological relevance of pulse detection for a single neuron is
questionable. In the case of channel noise, we come to a similar
conclusion. Therefore, we propose a feasible solution for a
neuronal population to overcome this predicament. We find that
subthreshold signal detection can be greatly enhanced with the
neuronal networks we propose. The phenomenon of intrinsic SR
induced by channel noise is also observed. We argue this SR may be
a strategy that neural systems would take to optimize their
detection ability for subthreshold signals.

Our paper is organized as follows. In Sec. II the stochastic
version of the Hodgkin-Huxley neuron model is presented. In Sec.
III, we focus on how the single neuron responses to subthreshold
transient input pulse. Phase plane analysis method is applied to
explain results presented. In Sec. IV, we present the simple
scenario for pulse detection and demonstrate that the detection
ability of a single neuron is limited. Then we introduce the
network that could reliably detect subthreshold pulses.
Discussions and conclusions are presented in Sec. V.

\section{MODELS}

\subsection{Deterministic Hodgkin-Huxley Model}

The conductance-based Hodgkin-Huxley (HH) neuron model provides a direct relationship between the microscopic properties of an ion channel and the macroscopic behaviors of a nerve membrane~\cite{Hodgkin-Huxley}. The membrane dynamics of the HH equations are given by
\begin{eqnarray}
C_{m} \frac{dV}{dt} &=& - \left( G_{K} (V-V_{K}^{rev}) + G_{Na} (V-V_{Na}^{rev}) + G_{L}(V-V_{L}) \right) +I,
\label{eq-1}
\end{eqnarray}
where $V$ is the membrane potential. $V_{K}^{rev}$ and $V_{Na}^{rev}$, $V_{L}$ are the reversal potentials of ($K$) potassium and ($Na$) sodium, the leakage currents, respectively. $G_{K}$, $G_{Na}$, and $G_{L}$ are the corresponding specific ion conductances. $C_{m}$ is the specific membrane capacitance, and $I$ is the current injected into this membrane patch. The conductance for potassium and sodium ion channels are given by
\begin{equation}
G_{K}(V,t) = \overline{g}_{K} n^{4}, \qquad G_{Na}(V,t) =
\overline{g}_{Na} m^{3}h, \label{eq-2}
\end{equation}
where $\overline{g}_{K}$ and $\overline{g}_{Na}$ are products of two factors: an individual channel conductance $\gamma_{K}$ and $\gamma_{Na}$ respectively, and the channel densities $\rho_{K}$ and $\rho_{Na}$ respectively. $\overline{g}_{K}$ and $\overline{g}_{Na}$ give the maximum conductance when all channels are open. The gating variables, $n$, $m$, and $h$, obey the following equations,
\begin{eqnarray}
\frac{d}{dt} n &=& \alpha_{n}(V)(1-n) - \beta_{n}(V)n, \nonumber \\
\frac{d}{dt} m &=& \alpha_{m}(V)(1-m)-\beta_{m}(V)m, \nonumber \\
\frac{d}{dt} h &=& \alpha_{h}(V)(1-h)-\beta_{h}(V)h,
\label{eq-3}
\end{eqnarray}
where $\alpha_{x}(V)$ and $\beta_{x}(V)$ ($x=n$, $m$, $h$) are voltage-dependent opening and closing rates and are given in Table~\ref{table-1} with the other parameters used in the following simulations.

\begin{table}[h]
\caption{Parameters and rate functions used in our Simulations.}
\begin{tabular*}{8.5cm}{@{\extracolsep{\fill}}lll}
\hline
$ C_{m}       $  & Specific membrane capacitance    & $1\mu F / cm^{2}$ \\
$V^{rev}_{K}  $  & Potassium reversal potential     & $ -77mV        $ \\
$ V^{rev}_{Na}$  & Sodium reversal potential        & $50mV$ \\
$ V_{L}       $  & Leakage reversal potential       & $-54.4mV$ \\
$ \gamma_{K}  $  & Potassium channel conductance    & $20pS$\\
$ \gamma_{Na} $  & Sodium channel conductance       & $20pS$\\
$ G_{L}       $  & Leakage conductance              & $0.3mS/cm^{2}$\\
$ \rho_{K}    $  & Potassium channel density        & $20/\mu m^{2}$\\
$ \rho_{Na}   $  & Sodium channel density           & $60/\mu m^{2}$\\
$ \alpha_{n}  $  & &$\frac{0.01(V+55)}{1-e^{-(V+55)/10}}~ ms^{-1}$\\
$\beta_{n}    $  & & $0.125e^{-(V+65)/80} ~ ms^{-1}$\\
$\alpha_{m}   $  & & $\frac{0.1(V+40)}{1-e^{-(V+40)/10}} ~ ms^{-1}$\\
$\beta_{m}    $  & & $4e^{-(V+65)/18}~ ms^{-1}$\\
$\alpha_{h}   $  & & $ 0.07e^{-(V+65)/20}~ ms^{-1}$\\
$\beta_{h}    $  & & $ \frac{1}{1+e^{-(V+35)/10}}~ ms^{-1}$\\
\hline
\end{tabular*}
\label{table-1}
\end{table}

\subsection{Stochastic Hodgkin-Huxley Model}

The deterministic HH model describes the average behaviors of a
larger number of ion channels. However, ion channels are random
devices, and for the limited number of channels, statistical
fluctuations play a role in neuronal dynamics~\cite{Abbott}. To treat the consequent fluctuations in
ion conductance, two kinds of methods are often employed.

One is the so-called Langevin method which characterizes channel noise with Gaussian white noise~\cite{Fox}. In this description, the voltage variables still obey Eqs. (\ref{eq-1}) and (\ref{eq-2}) but the gating variables are random quantities obeying the following stochastic differential equations,
\begin{eqnarray}
\frac{d}{dt}n &=& \alpha_{n}(V)(1-n) - \beta_{n}(V)n + \xi_{n}(t), \nonumber \\
\frac{d}{dt}m &=& \alpha_{m}(V)(1-m) - \beta_{m}(V)m + \xi_{m}(t), \nonumber \\
\frac{d}{dt}h &=& \alpha_{h}(V)(1-h) - \beta_{h}(V)h + \xi_{h}(t),
\label{eq-4}
\end{eqnarray}
where the variables $\xi_{n}(t)$, $\xi_{m}(t)$, and $\xi_{h}(t)$
denote Gaussian zero-mean white noise with
\begin{eqnarray}
\left< \xi_{n}(t) \xi_{n}(t') \right> &=& \frac{2}{N_{K}} \frac{\alpha_{n}(V)(1-n) - \beta_{n}(V)n}{2} \delta(t-t'), \nonumber \\
\left< \xi_{m}(t) \xi_{m}(t') \right> &=& \frac{2}{N_{Na}} \frac{\alpha_{m}(V)(1-m) - \beta_{m}(V)m}{2} \delta(t-t'), \nonumber\\
\left< \xi_{h}(t) \xi_{h}(t') \right> &=& \frac{2}{N_{Na}}
\frac{\alpha_{h}(V)(1-h) - \beta_{h}(V)h}{2} \delta(t-t'),
\label{eq-5}
\end{eqnarray}
where $N_{K}$ and $N_{Na}$ are the total number of $K^+$ and $Na^+$ channels. Note that in this description, a precondition is that $n$, $m$, and $h$ should be in the interval $[0,1]$. It has been argued that the Langevin method cannot reproduce accurate results (see \cite{Shangyou Zeng} for detail). However, it is still an effective method and widely used for its low computational cost. Additionally, the trajectory of the phase point prior to a spike entails major changes in the variables $V$ and $m$ but the variables $n$ and $h$ are practically unchanged during the same epoch~\cite{Tuckwell}. So, this recipe enables us to investigate system behaviors in the $V-m$ phase plane.

\begin{figure}[h]
\includegraphics[width=0.5\textwidth]{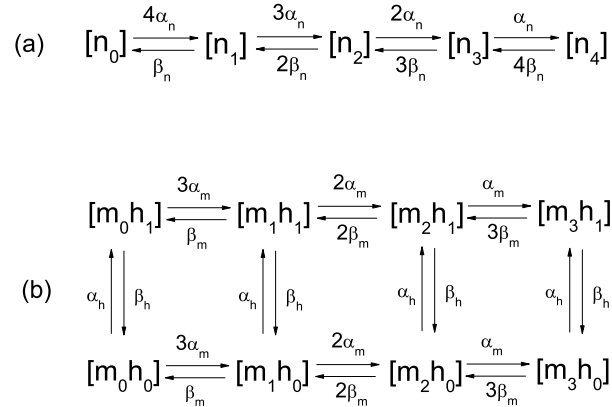}
\caption{Kinetic scheme for a stochastic potassium channel (a) and sodium channel (b). $n_{4}$ and $m_{3}h_{1}$ are open states, while the other states are non-conducting.}
\label{fig1}
\end{figure}

Another method is based on the assumption that the opening and
closing of each gate of the channel is a Markov process. With this
methods, the ion channel stochasticity is introduced by replacing
the stochastic equations by the explicit voltage-dependent
Markovian kinetic models for a single ion
channel~\cite{Schneidman,White2, Jung}. As shown in
Fig.~\ref{fig1}, the $K^{+}$ channels can exist in five different
states and switch between these states according to the voltage
dependence of the transition rates (identical to the original HH
rate functions). $n_{4}$ labels the single open state of the
$K^{+}$ channel. The $Na^{+}$ channel kinetic model has $8$
states, with only one open state $m_{3}h_{1}$. Thus the
voltage-dependent conductances for $K^+$ and $Na^+$ channels are
given by
\begin{equation}
G_{K}(V,t) = \gamma_{K} [n_{4}] / S, \qquad G_{Na}(V,t) = \gamma_{Na} [m_{3}h_{1}]/S,
\label{eq-6}
\end{equation}
where $\gamma_{K}$ and $\gamma_{Na}$ are defined as before, and $[n_{4}]$ refers to the number of open $K^+$ channels, $[m_{3}h_{1}]$ the number of open $Na^+$ channels, and $S$ the membrane area of the neuron.

The numbers of open $K^+$ and $Na^+$ channels at a special time $t$ is determined by the following formula: if the transition rate between state $A$ and state $B$ is $r$ and the number of channels in these states is denoted by $n_{A}$ and $n_{B}$, the probability that a channel switches within the time interval $(t, t + \Delta t)$ from state $A$ to $B$ is given by $p = r\Delta t$. Hence, for each time step, we determine $\Delta n_{AB}$, the number of channels that switch from $A$ to $B$, by choosing a random number from the following binomial distribution,
\begin{equation}
P\left( \Delta n_{AB} \right) = \left( \begin{array}{c} n_{A} \\ \Delta n_{AB} \end{array} \right) p^{\Delta n_{AB}} (1-p)^{(n_{A}- \Delta n_{AB})}.
\label{eq-7}
\end{equation}
Then we update $n_{A}$ with $n_{A}-\Delta n_{AB}$, and $n_{B}$ with $n_{B}+\Delta n_{AB}$. To ensure that the number of channels in each state is positive, starting at the beginning with the largest rate, we update these numbers sequentially, and so forth~\cite{Shangyou Zeng}.

The noisiness of a cluster of channels can be quantified by the coefficient of variation (CV) of the membrane current. Under assumptions of stationarity ($V$ is fixed ), $CV = \left\{ (1-p)/np \right\} ^{1/2}$, where $n$ is the number of channels and $p$ is the probability for each channel to be open. Thus the noisiness for a given population of voltage-gated channels is proportional to $n^{1/2}$~\cite{White2}. Accordingly, in this study, we introduce the membrane area $S$ as a control parameter of the channel noise level. Given ion channel density, the level of channel noise decreases with an increase in membrane area.

The numerical integrations of stochastic equations for both the occupation number method and the Langevin method are performed by using forward Euler integration with a step size $0.01 ms$. The parameters used in all simulations are listed in Table~\ref{table-1}. The occurrences of action potentials are determined by upward crossings of the membrane potential at a certain detection threshold $10 mV$ if it has previously crossed the reset value of $-50 mV$ from below.

\section{The response of SHH neuron to a subthreshold transient input pulse}

The signal detection of transient subthreshold input pulses has received increasing attention in recent years~\cite{Boris,Hasegawa,Ginzburg} (see \cite{Gregor} for more references). In our study of the response of a SHH neuron, the transient input pulses are set with width $\delta t=0.1 ms$ and strength $I_{0}=5 ~\mu A/cm^{2}$.

\begin{figure}[h]
\includegraphics[width=0.4\textwidth]{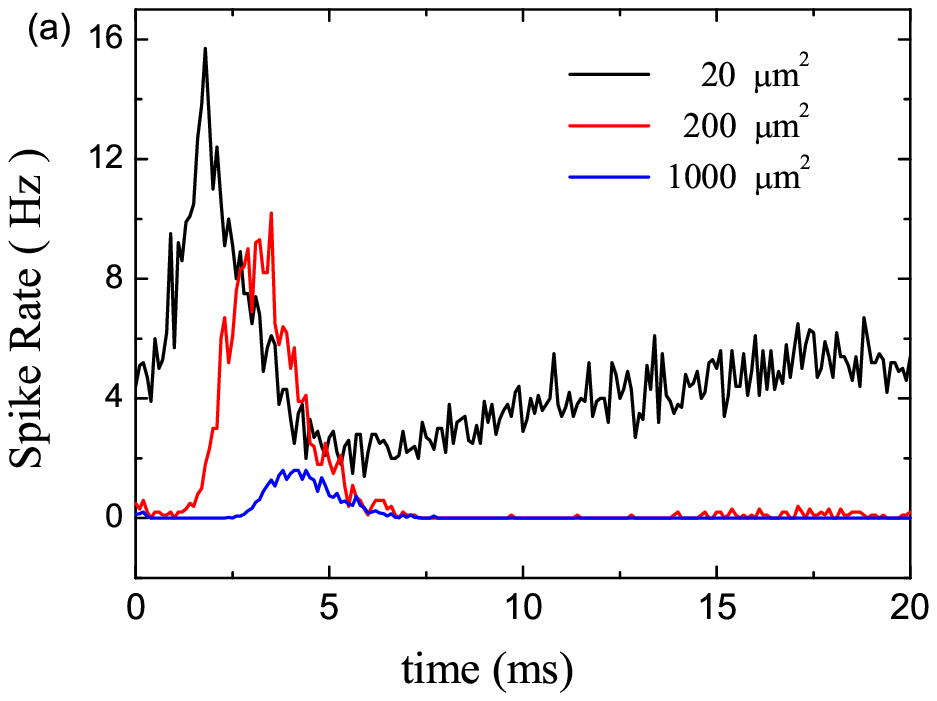}
\includegraphics[width=0.4\textwidth]{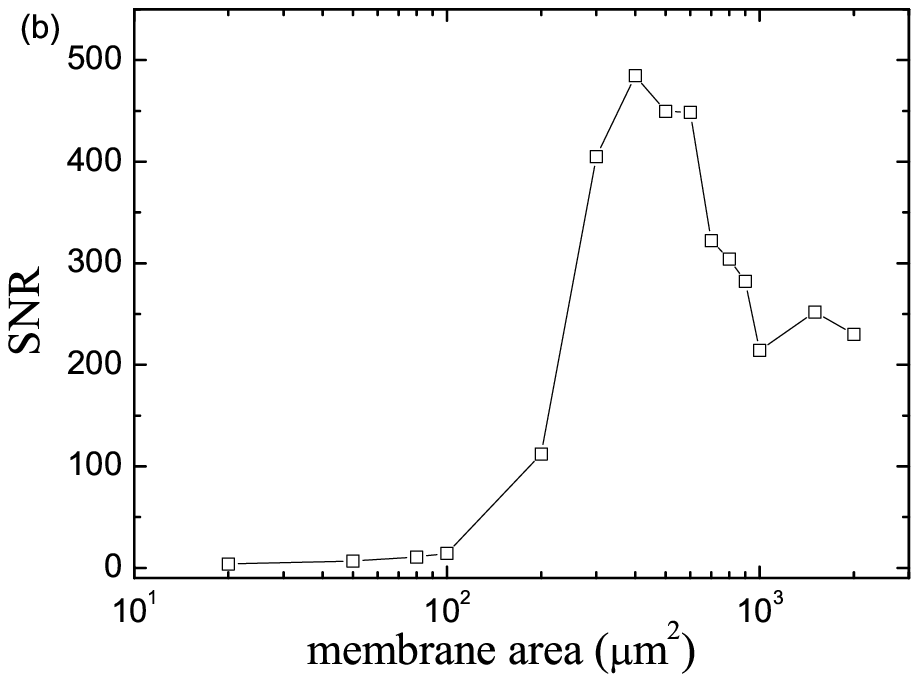}
\includegraphics[width=0.4\textwidth]{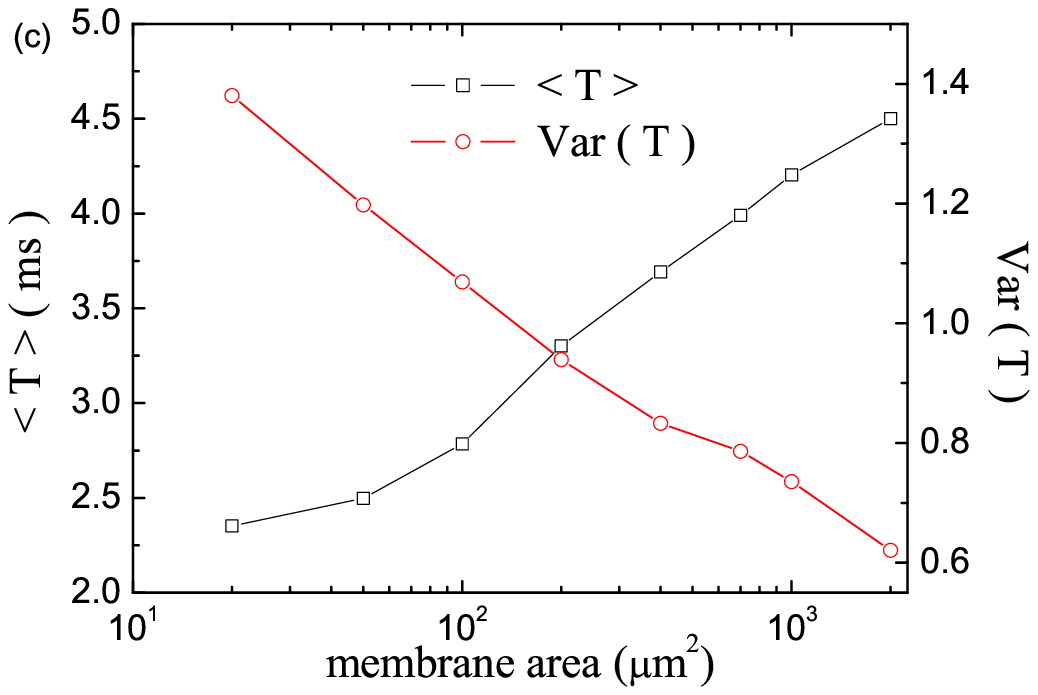}
\caption{(Color online) The response properties of a SHH neuron to subthreshold transient input pulses. (a) The PSTH of a SHH neuron with different membrane areas vs time. (b) SNR vs membrane area. (c) the mean response time and the variance of response time vs membrane area. }
\label{fig2}
\end{figure}

Fig.~\ref{fig2}(a) depicts the post-stimulus time histograms
(PSTHs) of a SHH neuron with a membrane area $S=20$, $200$, and
$1000 \mu m^2$, respectively. Each stimulus was repeated $5000$
times. The number of spikes observed in each bin (bin size
$=0.1ms$) is normalized by the total number of stimuli and by the
bin size. Thus, the PSTH gives the firing rate or the distribution
of the firing probability as a function of time~\cite{Svirskis2}.
Obviously, there exists a peak over the spontaneous firing level
in each curve and the peak lessens as the membrane area $S$
increases. The higher the peak, the more sensitive
 neuron responses are to stimuli, which are activated by channel
noise. The baselines show the average level of spontaneous firing
due to channel noise. With a higher baseline, the number of
spontaneous spikes increases. R. K. Adair has shown that the
firing rate of a neuron with channel noise can be reduced by
lowering the resting potential (Fig. 5 in Ref.~\cite{Adair}). In
our case, the transient input pulse temporally hold the resting
potential to a high state, thus gives a temporally higher firing
rate over the spontaneous one. As the membrane area increases,
since the fluctuations in membrane currents become smaller, the
firings in response to the subthreshold signals as well as the
noise-induced spontaneous firings are reduced, yielding reductions
in heights of both the peaks and the baselines. It is noted that
adjacent to the peak, there follows a time interval of about
$10ms$ during which the firing rate is below its average level. We
argue that this trough shape of the PSTH is due to refractoriness
of the neurons~\cite{Hodgkin-Huxley}. If in a certain time
interval the firing rate is higher than its average level, the
firing rate in the following time range will be reduced because
refractory effect prevents occurrence of the immediately following
firings. The time interval of $10ms$ is in accordance with
effective refractory period reported by other
researchers~\cite{Brown}.

To find the range in the membrane area which is more sensitive to
a pulse than channel noise perturbation, we define signal-to-noise
ratio (SNR) as the ratio of increased firing probability in
response to input pulses to the probability for spontaneous firing
in response to channel noise.~\cite{Svirskis2}. As shown in
Fig.~\ref{fig2}(b), when the membrane area is smaller than
$100~\mu m^{2}$, SNR remains very small. With increasing membrane
area, SNR increases rapidly and reaches its maximum at about $400
\mu m^{2}$. However, further increasing the membrane area leads to
a decreasing in SNR. This figure clearly demonstrates the
phenomenon of stochastic resonance. It is noted that as the
membrane area increases, both the peak and the baseline of PSTH
 is reduced, so the occurrence of SR for SNR curve is a result of trade-off between neuron's
sensitivity to subthreshold signals and rejection of spontaneous
firings.

Next, we investigate how the channel noise affects the response
time of neurons to subthreshold signals. It has been recently
proposed that the first spikes which occur in, for example,
cortical neurons, may contain information about a
stimulus~\cite{VanRullen}. Thus, determinacy in response time of a
neuron to signals is relevant to the information content, and how
it is affected by channel noise would be an important question to
explore~\cite{Tuckwell}. The PSTH analysis provides us a first
glimpse into it. The central positions of the PSTH peaks represent
the mean response time, and the widths of the peaks represents the
variances in response time. We see that as membrane area is
increased, the central position of PSTH peak moves rightward, and
 its width is reduced simultaneously[see Fig.~\ref{fig2}(a)]. This implies that
 as the membrane area increases, the mean response time increases but the variance in response time decreases.
 This is in consistent with the results obtained in the
 case of subthreshold inputs for external noise ~\cite{Tuckwell2}.
 From 5000 times repeated trials, we directly calculate the mean and variance of response time for different
 membrane area, and plot them in Fig.~\ref{fig2}(c). It seems that
 the increasing of $\left< T \right>$ as well as the decreasing of
$Var(T)$ is nearly in an exponential form.

\begin{figure*}[h]
\includegraphics[width=1.0\textwidth]{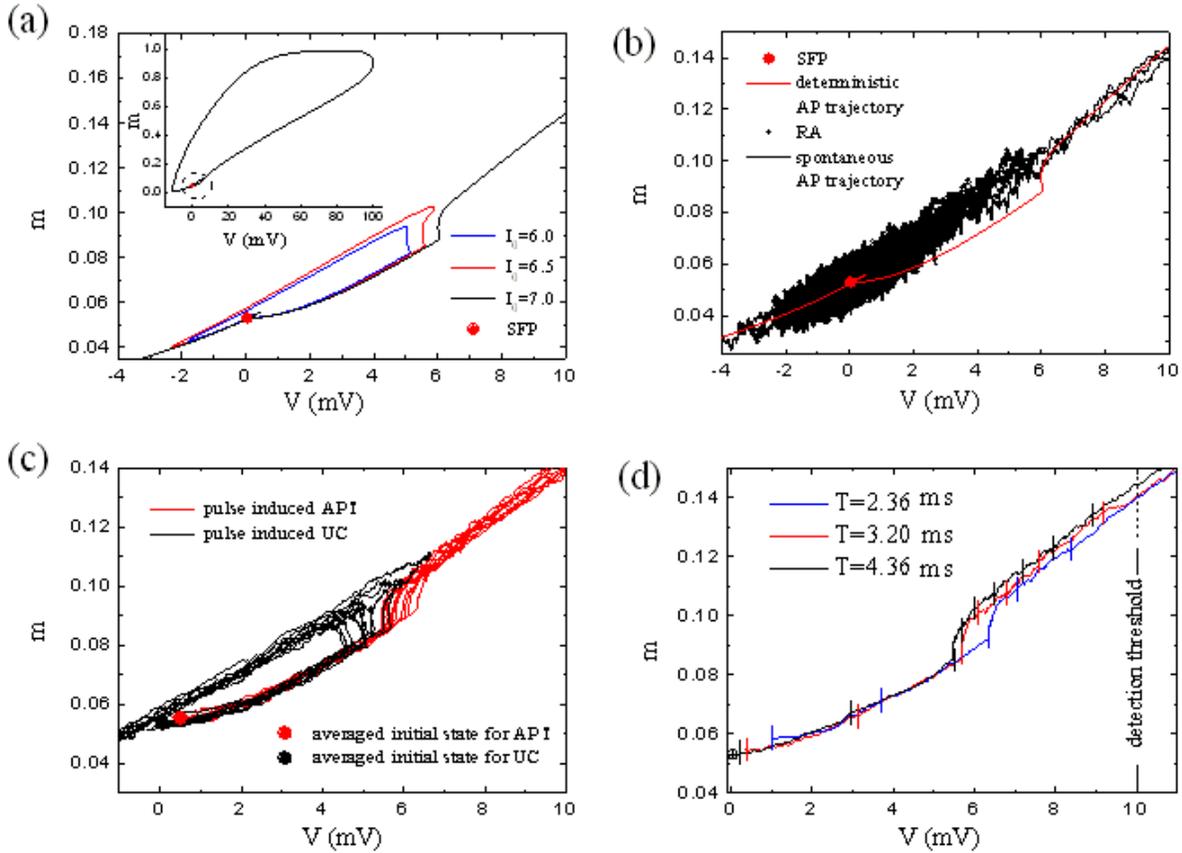}
\caption{(Color online) The phase plane analysis for the response properties of SHH
neuron with Langevin simulations. (a) The action potential
trajectory (APT) and the unstable circle of a noise-free HH neuron
in the phase plane. Inset: the overall APT for $I_{0}=7.0$ and the
position expanded in (a) (marked with dashed circle). (b) The stable
area and the spontaneous APT for membrane area $S=100 \mu m^{2}$.
(c) The noise-induced APT and the unstable circle (UC) for $S=1000
\mu m^{2}$. (d) Equitime labeling analysis of the response time to
different initial states.} \label{fig3}
\end{figure*}

\begin{figure*}[h]
\includegraphics[width=0.45\textwidth]{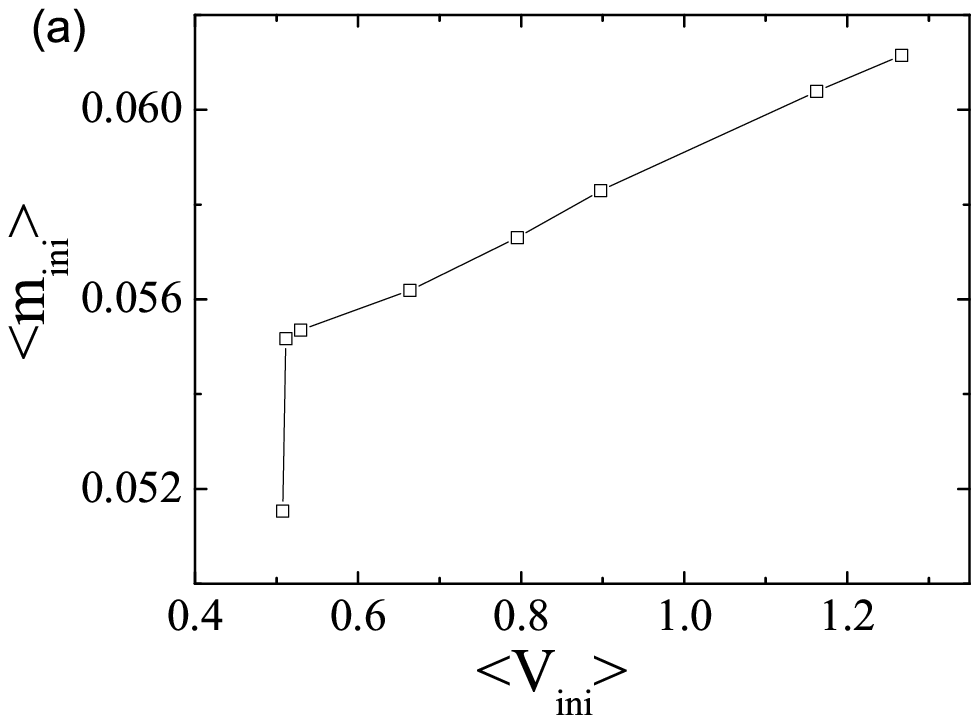}
\includegraphics[width=0.45\textwidth]{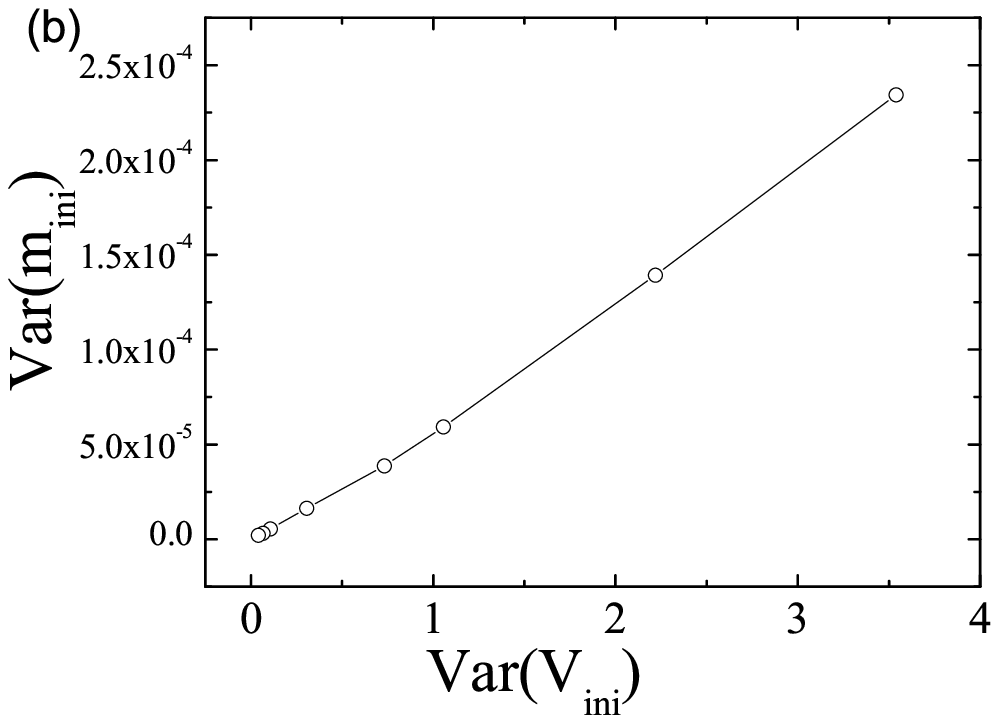}
\caption{The distribution of initial states for different membrane areas. (a) The averages of $V_{ini}$ and $m_{ini}$ for different membrane areas. (b) The variances of $V_{ini}$ and $m_{ini}$ for different membrane areas. The eight different dots, from top to bottom in each plot, correspond to membrane area $S=20$, $50$, $80$, $100$, $200$, $500$, $800$, $1000 ~\mu m^{2}$, respectively.}
\label{fig3b}
\end{figure*}
To investigate the dynamic mechanism of a SHH neuron responding to transient input pulses,
we performed a phase plane analysis with the Langevin simulation model described above.
 $\delta t$ and $I_{0}$ for the input pulses are set as $1ms$ and $6\mu A/cm^{2}$, respectively.
Fig.~\ref{fig3}(a) shows the stable fixed point (SFP), part of the action potential trajectory (APT),
and the unstable circles (UC) corresponding to different intensities of the input pulses in a noise-free HH model.
The whole APT for $I_{0}=7.0$ is demonstrated in the inset of Fig.~\ref{fig3}. Note that there exists
a threshold in this system. The larger the intensity of input pulses, the further the system will
be displaced from the SFP. If the displacement is larger than the threshold, an action potential is
generated and the system comes back to the SFP along the APT. Otherwise, the system evolves along a
relatively smaller unstable circle to the SFP [the color plots in Fig.~\ref{fig3}(a)], and merely causes
the subthreshold membrane potential fluctuations. When channel noise is involved, the system doesn't stay
on the original SFP but fluctuates around the vicinity of SFP,
which we call the resting area (RA) [the black area in Fig.~\ref{fig3}(b)].
Occasionally, the system runs across the threshold due to perturbations in channel noise,
  then the system will evolve along a stochastic AP trajectory and a spontaneous action potential occurs.
 In the case of smaller membrane areas or larger RA, it is easier for the system to reach the AP trajectory under noise perturbation and produces more spontaneous firings (not shown).

To understand how the system responds to the pulse input with a amount of noise, we traced ten
trajectories for pulses with firing and no firing, respectively. As shown in Fig.~\ref{fig3}(c),
in both cases, the system is displaced to an area around the threshold. Then after the stimulus is removed,
 the system jumps onto the APT to generate a spike or onto the unstable circle and returns back to the RA.
 This demonstrates the jumping is random. The more right the state is before jumping in phase plane,
 the greater possibility for it to jump onto the APT. The more left the state is before jumping,
 the more possible for the system to jump onto unstable circles. Since our discussions are limited to cases
 of a small amount of noise, noise cannot affect the length of the pulse displacement, and so the jumping area
 is determined by the initial state of the systems before the input pulse is applied
 (in our case, the state is described by two variables: $V$ and $m$; $n$ and $h$ are not considered).
 The initial states with larger $V$ and $m$ are more likely to lead to an action potential
 [see the averaged initial positions for firing and no firing plotted in Fig.~\ref{fig3}(c)].
 Therefore we conclude that the response of the single SHH neuron to input pulses is state-dependent.

We also investigated the temporal response of the SHH neuron in the $V-m$ phase plane.
 In Fig.~\ref{fig3}(d), the APTs with three different response times are traced and labeled
 with bars separated equally by $0.5ms$. The leftmost bars denote the time that the input
 pulse is applied, and the dashed line denotes the time that the spikes are detected.
 It shows that the system reaches a position closer to the detection threshold if the
 initial state is higher, and it will come into the outer side of the APT on which the
 system moves more quickly than the systems on the inside of the APT. As a result, this
system presents a shorter response time to the input pulse, and vice versa. We see that
the response time for a particular input pulse is dependent on the initial state of the system.
In addition, one could deduce that it is the variance of initial state that results in the variance of the response time.

Next, we investigated how the change of membrane area (ie., the channel noise level)
effects the distribution of initial state of the system, so that the response time
exhibits statistics as shown in (c) of Fig.~\ref{fig2}. The distributions of initial state for
different membrane areas are described by the average and variance of $V_{ini}$ and $m_{ini}$, which are calculated
from $2000$ firings in response to the pulses with Langevin simulation. It is seen from Fig.~\ref{fig3b} that as
 the membrane area increases, both the average and variance of $V_{ini}$ and $m_{ini}$ decreases.
 In other words, with the decreasing of channel noise, the distribution of initial states in the phase plane
 moves left-down to the lower value and becomes narrower. As we discussed above, lower initial state leads to
longer response time, and narrower distribution of initial states leads to smaller variance of the response time.
Therefore,  the average response time is prolonged and its variance is reduced if the membrane area is increased.

It is noted that because of large computational cost of our model,
it is difficult to obtain the statistical properties of response
time for larger membrane area. But through the phase plane
analysis we see that the most inner APT results in the maximal
response time, which is the up limit of the average response time.
As the membrane area increases to infinitely large, the average
response time will increase gradually to this maximal value. We
argue this maximal response time corresponds to the response time
of the noise-free HH neuron to the pulse of which the strength to
elicit a spike is minimal. In the deterministic HH model, for the
pulse input with the minimal strength of $I_{0}=6.92 ~\mu
A/cm^{2}$ and $\delta t=0.1 ms$, the maximal response time we
obtained is $5.87ms$, which matches the PSTH peaks' right edges
(see curves for $200 \mu m^{2}$ and $1000 \mu m^{2}$ in
Fig.~\ref{fig2}(a)). This time scale is important for the choosing
of the coincidence time window $\rho$ introduced in section IV. If
$\rho$ is much larger than the maximal response time the SHH
neuron could provide, spontaneous firings will be detected
together with the stimulated firings, thus reduce the accuracy of
detection. On the contrary, if $\rho$ is far less than it, some
stimulated firings will be ignored, so the efficiency of detection
is reduced.

\section{the performance of pulse detection }

\begin{figure}
\includegraphics[width=0.50\textwidth]{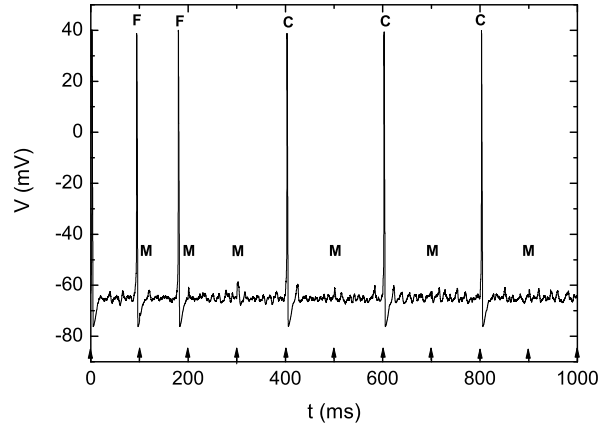}
\caption{Trace of the membrane potential of the SHH neuron for transient input pulses with width of $0.1ms$ and height of $5 ~\mu A/cm^{2}$. The time of occurrence for pulses is marked by arrowheads on the horizontal axis. $C$: the incidences that the pulse is correctly detected by the neuron; $M$: the incidences that the neuron does not respond to the pulse; $F$: the incidences that a spike occurs in the absence of a pulse.}
\label{fig4}
\end{figure}

Now, we consider the pulse detection task as a simple computation that a neural system can perform, in which we evaluate the performance of a single SHH neuron as well as a SHH neuron assembly.

The input $I_{stim}$ is modeled as a serial narrow rectangular
current pulse with width $\delta t=0.1 ms$ and strength $I_{0}=5
~\mu A/cm^{2}$ (see Fig.~\ref{fig4}). The input pulse train (the
arrowheads on the horizontal axis) is regular with a large time
interval $\Delta T = 100ms$. Compared to the membrane time constant,
the preceding pulse has no significant influence on the following
one. In such an arrangement, as has been discussed above, the SHH
neuron has three different responses (marked with different capital
letters respectively in Fig.~\ref{fig4}) to the pulse train which
consists of $n$ equidistant pulses:

\begin{enumerate}

\item [(1)] $C$ : The neuron generates an action potential immediately (within the time rang of $5ms$) after a pulse is presented,
which signifies successful detection of the pulse. We define $P_{C}$ as the fraction of correctly detected pulses,
which is the total number of correctly detected pulses, divided by the total number of input pulses.

\item [(2)] $M$ : The neuron fails to fire a spike immediately (within the time rang of $5ms$) after the pulse is presented. If we define $P_{M}$ as the fraction of missed pulses, then we have
$P_{M}=1-P_{C}$.

\item [(3)] $F$ : The neuron fires a spike in the absence of an input pulse (a false positive event). A deterministic HH neuron cannot fire spikes when the stimulus is not applied or if it is below the threshold. However, in the case of channel noise, stochastic effects give rise to spontaneous spiking. To describe the effect of those spontaneous firing spikes on subthreshold pulse detection, we denote $P_{F}$ as the total number of false positive events divided by the total number of input pulses. Note that $P_{F}$ can easily exceed $1$.

\end{enumerate}

In order to quantify the neuron's response to the pulse train, we define the total error $Q$ for the pulse detection,
\begin{equation}
Q = P_{M} + P_{F}. \label{eq-8}
\end{equation}
For a longer interval $\Delta T$ with fixed $n$, the false positive events are more likely to occur and the total error grows with increasing $\Delta T$.

\begin{figure}[h]
\includegraphics[width=0.5\textwidth]{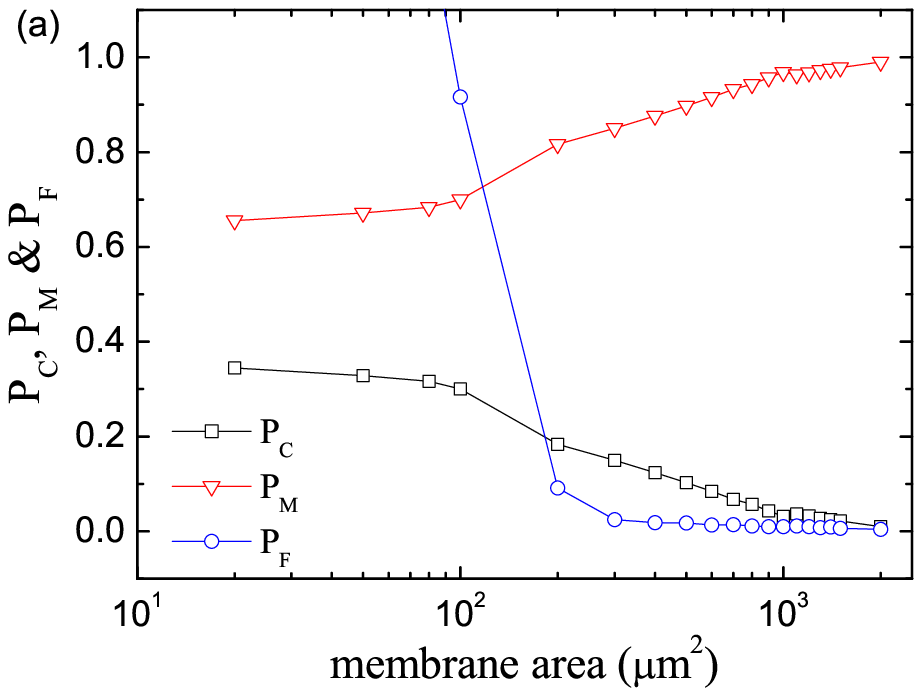}
\includegraphics[width=0.5\textwidth]{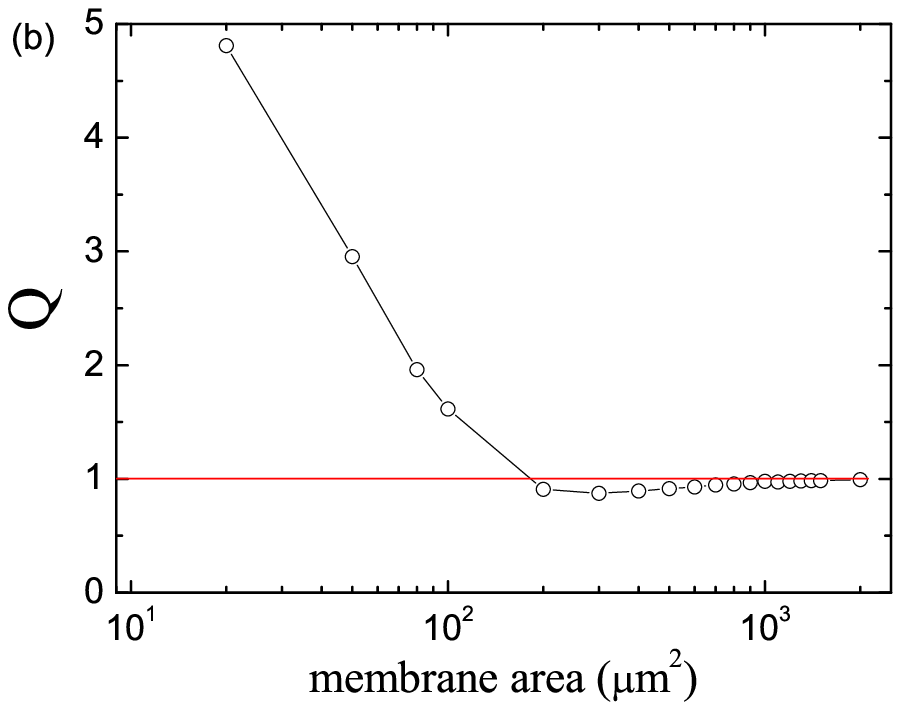}
\caption {(Corlor online) The performance of a single neuron in a subthreshold pulse detection task. (a) $P_{C}$, $P_{M}$, $P_{F}$, and (b) total error $Q$ as a function of membrane area.}
\label{fig5}
\end{figure}

Fig.~\ref{fig5}(a) shows $P_{C}$, $P_{M}$, and $P_{F}$ as a
function of the membrane area under the above-mentioned scenario.
According to the above PSTH analysis , when the membrane area is
rather large, though the system could be displaced by subthreshold
pulses to near the threshold, the channel noise is small and can
hardly trigger firings, thus $P_{C}$ is very small and $P_{M}
\approx 1$. The firings triggered by noise alone is even less, so
the $P_{F}$ is smaller than $P_{C}$. When the membrane area is
small, the channel noise is remarkable, giving large $P_{C}$.
Meanwhile, due to the high rate of spontaneous firings, $P_{F}$ is
even larger than $P_{C}$. So, with increasing membrane area, both
$P_{C}$ and $P_{F}$ decreases, but $P_{F}$ drops more quickly than
$P_{C}$. When the membrane area is larger than about $180 \mu
m^{2}$, $P_{C}$  becomes larger than $P_{F}$.

The total error $Q$ as a function of the membrane area is also
plotted in Fig.~\ref{fig5}(b). As the membrane area increases, due
to the rapid decline of $P_{F}$, the total error $Q$ drops
rapidly. Then, with further increase in membrane area, $Q$
increases and approaches $1$ for the major contribution from the
fraction of missed pulses. Because $Q$ is basically the summation
of ascending $P_{M}$ curve and descending $P_{F}$ curve, one can
expect a minimal value for it. The minimal value of $Q$ is
$0.8746$ at $S = 300 \mu m^{2}$. With this optimal membrane area,
we see that the neuron achieves balance between detecting pulse
input and suppressing spontaneous firings.

It should be noted that the positions for $P_{C}$, $P_{M}$ and
$P_{F}$ curves are dependent on the strength of pulse input
$I_{0}$, or the interpulse interval $\Delta T$. Changing the pulse
strength would change the position of $P_{C}$ curve, thus the
position of $P_{M}$ in Fig.~\ref{fig5}(a). In particular, smaller
pulse strength leads to lower position of $P_{C}$, thus higher
position of $P_{M}$(see Fig.~7 of Ref.\cite{Gregor} or Fig.~6 of
Ref.\cite{Adair}). As discussed above, the pulse induced high
firing rate would reduce the spontaneous firing rate through
refractoriness within the following $10 ms$. If $\Delta T$ is
large compared to refractory period of SHH neuron, this reduction
in spontaneous firings is negligible, so the pulse strength would
not affect the position of $P_{F}$ curve. On the contrary, by
their definition, $P_{F}$, rather than $P_{C}$ and $P_{M}$, is
greatly dependent on $\Delta T$. Whatsoever, neither pulse
strength nor interpulse interval will not change the overall shape
of both $P_{C}$ and $P_{F}$ curves. Since the minimal $Q$ is
basically the result of summation of ascending $P_{M}$ curve and
descending $P_{F}$ curve, we argue that there is always a minimal
value for $Q$, and the optimal membrane area for $Q$ differs for
different input pulse strength or interpulse interval.

We see that the single neuron has limited capacity for
subthreshold signal detection. The channel noise is basically a
zero-mean noise, which means the probability for a subthreshold
pulse gets enhanced by a positive fluctuation is equal to the
probability that it is further suppressed by its negative
counterpart. In more detail, the response of the SHH neuron to
input current pulses is state-dependent(Fig.~\ref{fig3}(c)).
Channel noise perturbations enable the system, with equal chance,
to be in a high state that the neuron is more likely to fire a
spike after pulse is applied, or in a low state that no fires
occur. As a result, $P_{C}$ could never exceed 0.5, and the total
error for a single neuron is always larger than 0.5. Indeed, we
found in Fig.4 of Ref.~\cite{Adair} that the spike efficiency for
subthreshold voltage impulses never exceed $50\%$ and the same
conclusion was made in Ref.~\cite{Gregor} for external noise. So
we see that theoretically, it is unlikely to utilize channel noise
to reliably detect subthreshold signals with single neuron.
However, in reality, the neuron assembly works in real neural
systems rather than in a single neuron. In general, neurons work
cooperatively through synaptic coupling. What's more, among
various spatiotemporal spike patterns in the neural system,
synchronous firing has been most extensively studied both
experimentally and theoretically~\cite{Kitajima,Ward}. Believing
that neuronal synchronous firing is critical for transmitting
sensory information, many investigators have suggested that a
major function of cortical neurons is to detect coincident events
among their presynaptic inputs (see~\cite{Roy} for more
references). Based on this fact, we proposed a neuronal network
that can greatly enhance the detection ability of the pulse.

\begin{figure}
\includegraphics[width=0.5\textwidth]{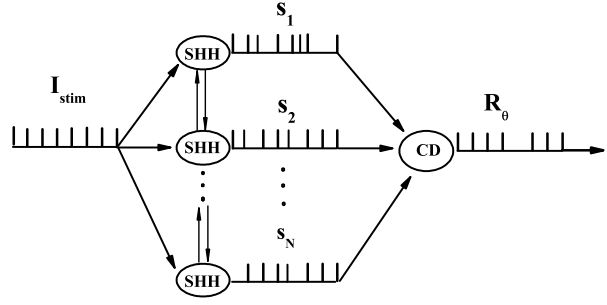}
\caption{A schematic diagram of pulse detection with multiple neurons. $I_{stim}$ is the input pulse train. $\textbf{S}_{i}$ is the output spike train of the $i$th SHH neuron for $i=1,2,\ldots,N$. $CD$ is the coincidence detector neuron and $\textbf{R}_{\theta}$ is its output spike train with synchronous firing detecting threshold $\theta$. Here we demonstrate the case of $\theta=2$.}
\label{fig6}
\end{figure}

As shown in Fig.~\ref{fig6}, the front layer of the network is
composed of globally coupled identical neurons with channel noise.
The coupling term has the form of an additional current $I_{couple}$
added to the equation for the membrane potential (see
Eq.~\ref{eq-1}). For the $i$th neuron, it takes the form
\begin{equation}
I_{couple} = \frac{\varepsilon}{N} \sum _{j=1}^N (V_{j}-V_{i}),
\label{eq-9}
\end{equation}
where $\varepsilon$ is the coupling strength and $V_{i}$ is the
membrane potential of the $i$th neuron for $i=1,\ldots,N$. In our
simulations, neurons are weakly coupled, $\varepsilon=0.005$. And
the membrane area of each SHH neuron is set as $S = 200 \mu m^{2}$.
Here we chose this value for the membrane area rather than the
optimal one so that $P_{C}$ of a single neuron is relatively large.
Thus fewer neurons are needed in our network and the computational
cost is consequently reduced. Each SHH neuron in the network
receives the same subthreshold pulse train as in the single neuron
case. The output spike trains of those neurons $S_{i}$
($i=1,2,\ldots,N$) are taken as the input of a so-called coincidence
detector (CD) neuron. In neural reality, coincidence detection
requires complex cellular mechanisms~\cite{Edwards,Stuart}. For
simplicity, here we use a phenomenological CD neuron model. The CD
neuron is excited when it detects spikes from more than $\theta$
neurons within a coincidence time window $\rho$ ($=5~ms$, see
discussion Sec. V). In other words, $\theta$ denotes the detection
threshold of the CD neuron. After firing, the CD neuron enters a
refractory period of $5~ms$. Obviously, given the input spike trains
$S_{i}$, the output spike train $R_{\theta}$ of the CD neuron is
determined by its threshold $\theta$. We also define
$P_{C}^{~\theta}$ and $P_{F}^{~\theta}$ as the fraction of correct
detection and false reporting in the network with the CD threshold
$\theta$, respectively. Similarly, $Q^{~\theta}$ is defined as the
total error of the network with the CD threshold $\theta$.

\begin{figure}[h]
\includegraphics[width=0.45\textwidth]{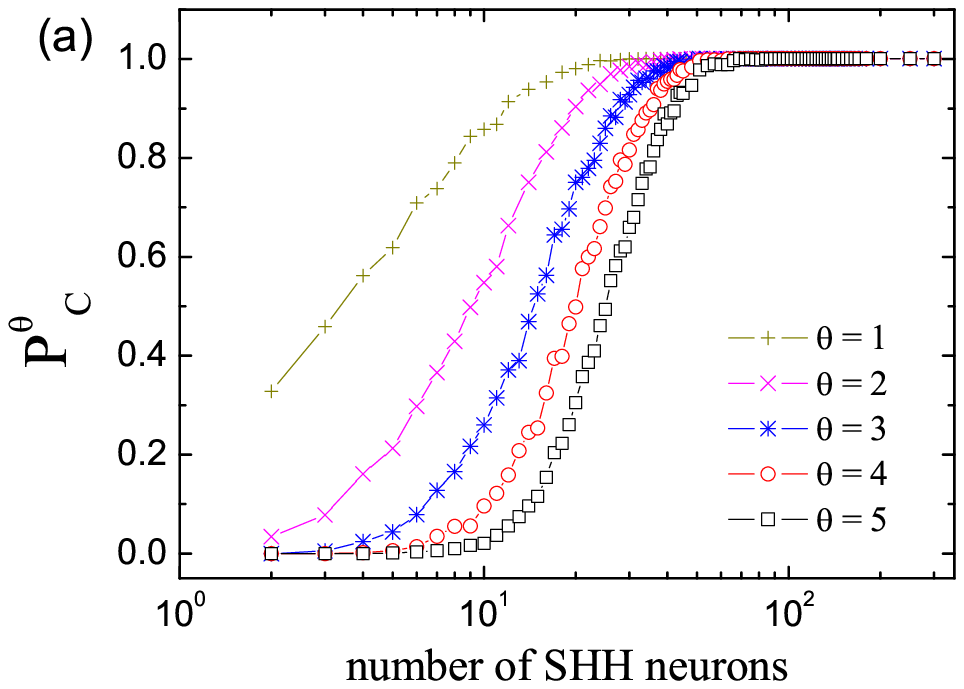}
\includegraphics[width=0.45\textwidth]{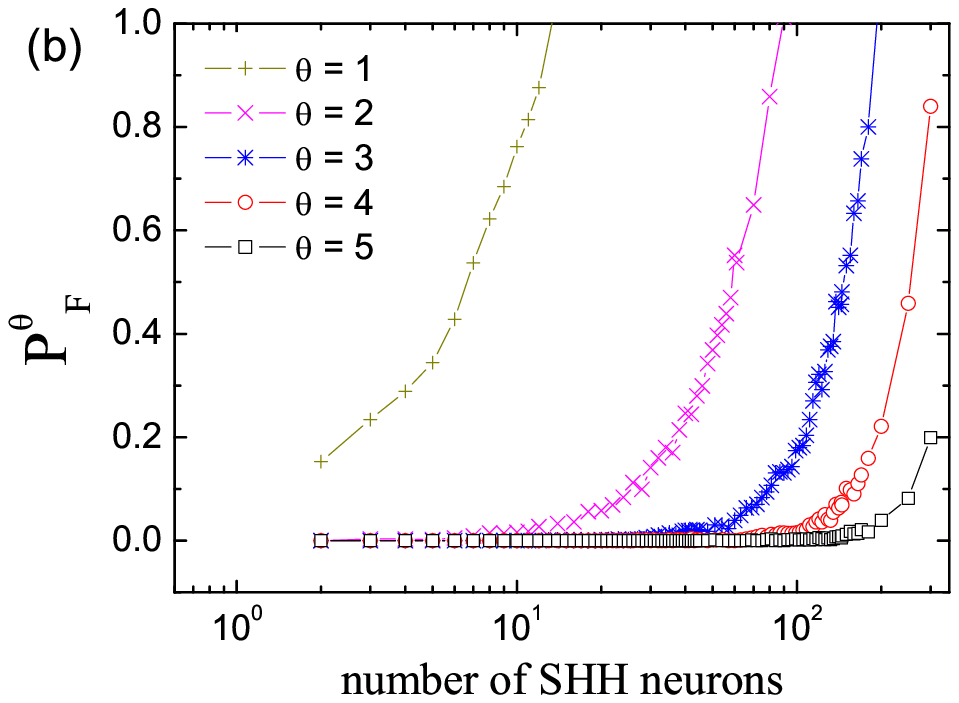}
\includegraphics[width=0.45\textwidth]{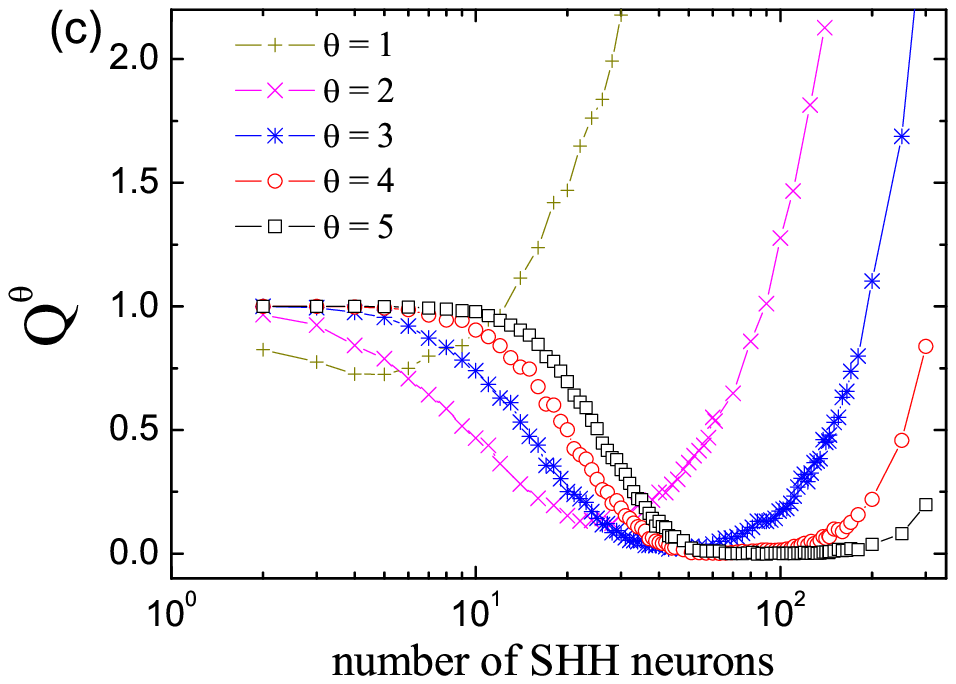}
\caption{(Color online) Detection of subthreshold signals with the neuronal
network for the CD threshold $\theta =1,2,...,5$. (a)
$P_{C}^{~\theta}$, (b) $P_{F}^{~\theta}$, and (c) $ Q^{~\theta}$
as a function of number of SHH neurons.} \label{fig7}
\end{figure}

\begin{table}[h]
\caption{The performance of the network with optimal sizes for
different CD threshold $\theta$.}
\begin{tabular*}{9.5cm}{c|ccccccc} \hline\hline
$\theta $ &~~1     &~~2    &~~3   &~~4  &~~5  &~~6\\
\hline
$N_{opt}$ &~~5     &~~22   &~~42   &~~$62\sim65$  &~~$69\sim91$ &~~$78 \sim 158$\\
$Q_{min}$ &~~0.725 &~~0.132&~~0.025   &~~0  &~~0  &~~0 \\
$P^{~\theta}_{C}$ &~~0.619 &~~0.937     &~~0.998 &~~1.0   &~~1.0  &~~1.0\\
\hline \hline
\end{tabular*}
\label{table-2}
\end{table}

 Fig.~\ref{fig7}(a) shows $P_{C}^{~\theta}$ as a function
of the number of SHH neurons $N$ for $\theta=1, 2, 3$. As the number
of neurons $N$ increases, all $P_{C}^{~\theta}$ increase quickly to
$P^{~\theta}_{C}=1$. For larger $\theta$, the increase of
$P_{C}^{~\theta}$ becomes slower and requires more neurons to
achieve the successful state $P^{~\theta}_{C}=1$. However, the
enhancement of $P_{C}^{~\theta}$ is at the cost of unexpected
improvement in $P_{F}^{~\theta}$. As shown in Fig.~\ref{fig7}(b),
with increasing $N$, $P_{F}^{~\theta}$ is also improved. Note that
$P_{F}^{~\theta}$ is able to exceed $1$. Comparing
Fig.~\ref{fig7}(a) with (b), it is obvious that, though both
$P_{C}^{~\theta}$ and $P_{F}^{~\theta}$ increase with an increased
number of neurons, comparing to $P^{~\theta}_C$, the increasing of
$P^{~\theta}_F$ is always delayed. Thus, in the case of a small $N$,
the correct detection will not be greatly enhanced though
$P_{F}^{~\theta}$ is low. Whereas for large $N$, one can obtain
better performance for correct detection but a cost of a higher
$P_{F}^{~\theta}$. Therefore, we expect to find an optimal $N$ to
achieve the best performances for signal detection.

Fig.~\ref{fig7} (c) displays the total error $Q^{~\theta}$ as a
function of the number of neurons $N$ for different $\theta$.
Clearly, the minimal total error $Q_{min}$ or the resonance
behavior appears at the network level. For different $\theta$,
there exist different optimal numbers of SHH neurons $N_{opt}$
where the performance of pulse detection is at its best. As shown
in Table~\ref{table-2}, with increasing $\theta$, $Q_{min}$
decreases while the corresponding $N_{opt}$ increases.
Simultaneously, $P^{~\theta}_{C}$ corresponding to the $N_{opt}$
increases. If $\theta$ is large enough, the $Q_{min}$ becomes
nearly zero ($< 0.001$) in a wide range of SHH neuron numbers.
Theoretically, by further enhancing the detection threshold and
involving more neurons, the zero value of $Q_{min}$  could appear
in a wider range of SHH neuron numbers.

\begin{figure}[h]
\includegraphics[width=0.45\textwidth]{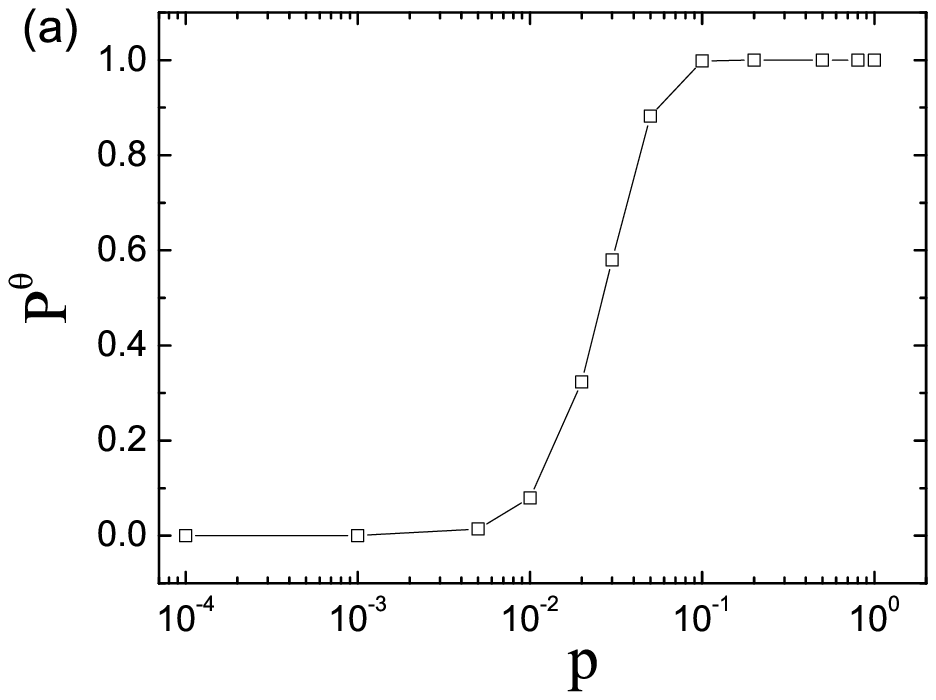}
\includegraphics[width=0.45\textwidth]{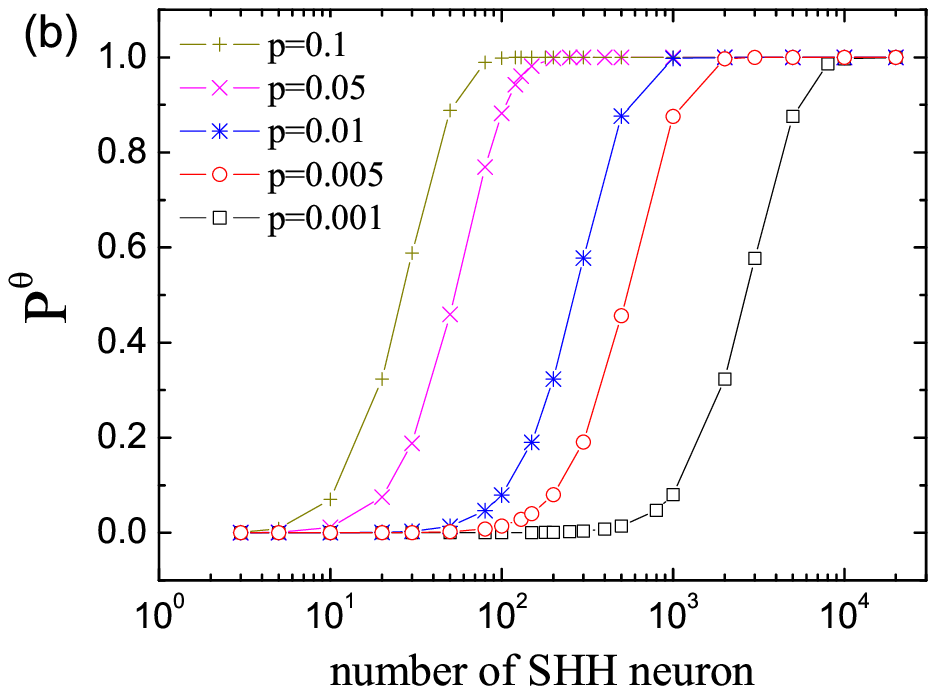}
\includegraphics[width=0.45\textwidth]{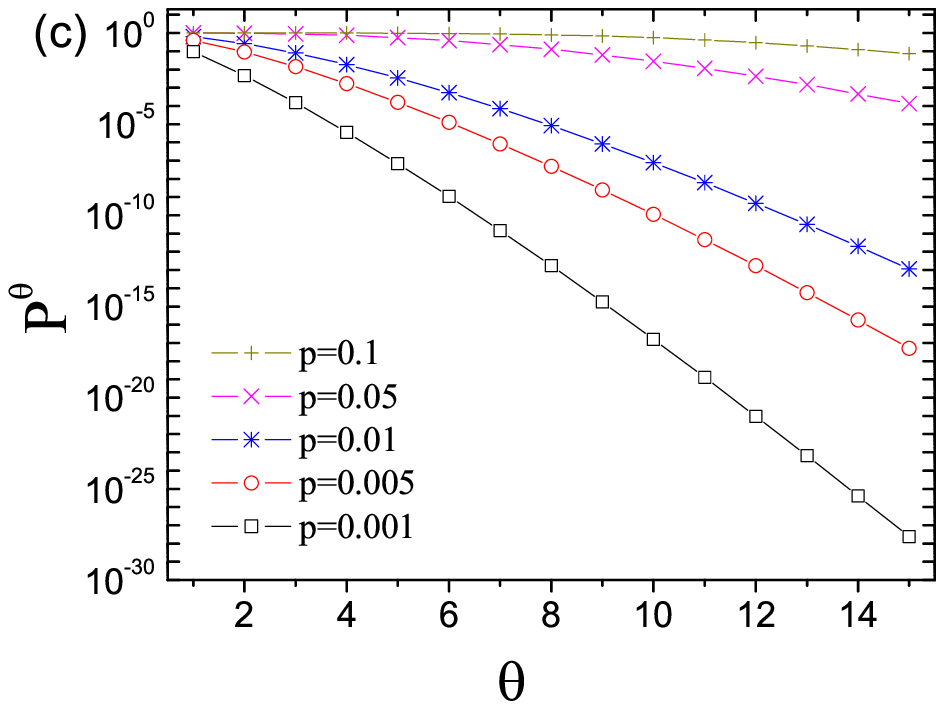}
\caption{(Corlor online) The syn-firing probability $P^{~\theta}$ of the network
calculated from Eq.~\ref{eq-10}. (a) $P^{~\theta}$ as a function
of $p$, the firing probability of each SHH neurons. $\theta=3$,
$N=100$. (b) $P^{~\theta}$ as a function of number of SHH neurons
for different $p$. $\theta=3$. (c) $ P^{~\theta}$ as a function of
$\theta$ for different firing probability of each SHH neurons $p$.
} \label{fig9}
\end{figure}

  We define syn-firing probability $P^{~\theta}$ as the probability
 that $\theta$ or more than $\theta$ SHH neurons fire in a time interval.
 Supposing the firing probability of each independent SHH neuron (ignoring the couplings between them)
  in a time interval is
 $p$, then syn-firing probability $P^{~\theta}$ in this time interval
 is described by cumulative distribution function for a binomial
 distribution, i.e.,

\begin{equation}
P^{~\theta}= \sum _{\alpha =\theta}^N
\frac{N!}{\alpha!(N-\alpha)!}p^{~\alpha}(1-p)^{N-\alpha},
 \label{eq-10}
\end{equation}

where $p^{~\alpha}(1-p)^{N-\alpha}$ is the probability that only
$\alpha$ neurons fire at the same time, and
$C^{N}_{\alpha}=\frac{N!}{\alpha!(N-\alpha)!}$ is the number of
ways of picking $\alpha$ neurons from population $N$.  So $\sum
_{\alpha =\theta}^N \frac{N!}{\alpha!(N-\alpha)!}$ is the total
number of ways of selecting $\theta$ or more than $\theta$ neuron
out of population $N$. Then the firing probability for CD neuron
with threshold $\theta$ and refractoriness is written as
\begin{equation}
P^{~\theta}_{CD}=
P^{~\theta}-(influence~~of~~past~~perturbations),
 \label{eq-11}
\end{equation}
where the last term represents the suppression of past firings on
present firing probability through refractoriness, which is
proportional to the probability for past firing events
~\cite{Herrmann}.  This term can be cancelled because it often
acts as small perturbations, and does not bring qualitative
changes to firing probabilities of CD neuron statistically. So in
the following analysis, for simplicity, we take $P^{~\theta}_{CD}
\approx P^{~\theta}$.

From Fig.~\ref{fig9}(a) we see that $P^{~\theta}$, thus $P_{CD}^{~\theta}$ increases with increasing  $p$, the firing probability of each neuron. So when pulses are applied to the SHH neuron,  the firing probability of CD neuron is larger than that caused by channel noise alone, because the firing probability
 of each SHH neuron is enhanced. By increasing the number of SHH
 neurons, no matter $p$ is large or small, $P^{~\theta}$, thus $P_{CD}^{~\theta}$
 increases(Fig.~\ref{fig9}(b)). Since as in the single neuron case,  $P_{C}^{~\theta}$ is proportional
 to $P_{CD}^{~\theta}$ in response to subthreshold pulses,
 and $P_{F}^{~\theta}$ is proportional to $P_{CD}^{~\theta}$ induced by channel noise
 alone.  $P_{C}^{~\theta}$ and $P_{F}^{~\theta}$ rise, but $P_{M}^{~\theta}$ declines (not shown) as the number of
 SHH neurons increases. Therefore, as a result of summation of
 declining $P_{M}^{~\theta}$ curve and ascending $P_{F}^{~\theta}$ curve, a minimum of the total error
 $Q^{~\theta}$ is warranted.

 As shown in Fig.~\ref{fig9}(c), with the increasing of $\theta$, $P^{~\theta}$ drops, and for small $p$, $P^{~\theta}$
 drops more quickly. So both $P_{C}^{~\theta}$ and $P_{F}^{~\theta}$ curves move right-down
 in Fig.~\ref{fig7} (a) and (b), and $P_{M}^{~\theta}$ curves move right-up(not shown). Thus, $ Q_{~\min}$ curves
 would move rightward in Fig.~\ref{fig7}(c). Since for single SHH neuron, its firing probability is low when pulse inputs are absent,
 $P_{F}^{~\theta}$ curves move more rapid than the $P_{M}^{~\theta}$ curves. As a result, the $Q^{~\theta}$ curves
 become wider, and move also downward in Fig.~\ref{fig7}(c) as $\theta$ increases. So we see that the drop of $ Q_{~\min}$ is warranted and
$ Q_{~\min}=0$ is expected to achieve in a wide range of SHH
neuron number when $\theta$ is large enough.

\section{Discussion and Conclusion}

In this paper, we used the stochastic version of Hodgkin-Huxley neuron model in which channel noise is the only source of noise, and discussed the possibility of detecting subthreshold signals with channel noise.

First, we studied the response property of the single SHH neuron
to the subthreshold transient input pulses. The main result is
that the SHH neuron fires spikes with a higher rate over its
average level in response to a subthreshold stimulus. The average
response time decreases but its variance increases as the channel
noise amplitude increases (or with decreasing membrane area). We
further found the existence of an up limit for the average
response time. From phase plane analysis we see that this up limit
should be predictable for threshold systems with any zero-mean
noise, as the noisiness decreases. This results means the response
time is very sensitive to the membrane area, because a small
decreasing of membrane area would lead to remarkable decreasing in
mean response time and increasing in its variance.

Adair has demonstrated the stochastic resonance in ion channels as the output response (in the  probability of action potential spikes, which is equivalent to $P_{C}$ in our paper.) from small input potential pulses across the cell membrane is
increased by added noise, but falls off when the input noise
becomes large. However, to evaluate the reliability of
subthreshold signal detection, one must consider not only the
response to subthreshold signals but also the spontaneous firings,
because from the standpoint of a neuron, those two kinds of output
make no difference. In this paper, we endowed the SHH with a
simple pulse detection scenario and calculated the total error
$Q$.  We found that a minimal $Q$ and the corresponding optimal
membrane area (noise). So we argue that to maximize the detection
ability, the strategy a neuron should take is balancing between
response to pulses and rejecting spontaneous firings, rather than
improving the response to pulses alone with optimal noise as Adair
demonstrated. As we argued, the first strategy allow to achieve
the minimal $Q$ for different pulse strengths, unlike the second
one with which is in effect only for large pulse strength (see Fig. 6 in Ref.~\cite{Adair}). However, even with the first strategy, we
found the detection ability of a single neuron is non-credible
because $Q$ cannot be larger than 0.5. Though the results are
obtained with channel noise, we argue the conclusion should be
general for any zero-mean noise.

The current SHH model is only an approximation to a much more
complex reality. For example, it has presumed that the channel
dynamics are Markov chain process, they act independently, and the
gating currents related to the movement of gating charges are
negligible. However, those presumptions are not always tenable.
For example, Schmid has shown that the gating currents drastically
reduce the spontaneous spiking rate if the membrane area is
sufficiently large~\cite{Schmid2}. So our results should be
reinvestigated with consideration of those factors. But we think
those factors would not bring qualitative changes to our results,
thus the general conclusions still hold.

We then investigated the subthreshold signal detection in a
neuronal network that concerns a coincidence detection neuron. We
found by enhancing  coincidence detection threshold and increasing
the SHH neurons, the detection ability is greatly improved. It
suggests that channel noise may play a role in information
processing in the neural network level. In addition, corresponding
to different coincidence detection thresholds, there exist an
optimal number of neurons at which the total error is at its
minimum. We have seen that this is also the result of balancing
between responding to pulses and rejecting spontaneous firings.
Since this so-called double-system-size resonance phenomenon has
been rarely reported~\cite{Maosheng Wang}, our work provides an
example of such an observation. In particular, with sufficiently
large coincidence detection threshold, the total error is zero in
a wide range of SHH neuron number, which means the detection
ability of this network could be robust against variance of neuron
number caused by cell production and death.

We have shown that the reliable detection of subthreshold signals
 with the network is predictable with probability theory, as long as
 each front layer neuron exhibits higher firing probability in response to signals than that induced by noise.
 Weak signal detection is also important in practice. For example,
mobile communications dictates the use of low power detection to
prolong the battery life. So our work suggests a possible way to
design reliable stochastic resonance detector for weak
signals~\cite{Saha}.

\section*{Acknowledgements}
We really appreciate two anonymous referees for their very constructive and helpful suggestions. This work was supported by the National Natural Science Foundation of China under Grant No. $10305005$ and by the Fundamental Research Fund for Physics and Mathematic of Lanzhou University.

\end{document}